\documentclass[pra,longbibliography,amsmath,amssymb]{revtex4-2}
\usepackage{graphicx}
\usepackage{dcolumn}
\usepackage{bm}
\usepackage{graphics}
\usepackage{slashed}
\usepackage{natbib}
\usepackage{amsthm}
\usepackage{url}
\usepackage[usenames,dvipsnames,svgnames,table]{xcolor}
\usepackage[T1]{fontenc}
\DeclareFontFamily{T1}{calligra}{}
\DeclareFontShape{T1}{calligra}{m}{n}{<->s*[1.44]callig15}{}
\DeclareMathAlphabet\mathcalligra   {T1}{calligra} {m} {n}
\DeclareMathAlphabet\mathzapf       {T1}{pzc} {mb} {it}
\DeclareMathAlphabet\mathchorus     {T1}{qzc} {m} {n}
\DeclareMathAlphabet\mathrsfso      {U}{rsfso}{m}{n}

\makeatletter
\newcommand{\tpitchfork}{%
  \vbox{
    \baselineskip\z@skip
    \lineskip-.52ex
    \lineskiplimit\maxdimen
    \m@th
    \ialign{##\crcr\hidewidth\smash{$-$}\hidewidth\crcr$\pitchfork$\crcr}
  }%
}
\makeatother

\newtheorem{theo}{Theorem}

\newtheorem{lemma}{Lemma}

\newtheorem{cor}{Corollary}
 
\begin{document}

\preprint{APS/123-QED}

\title{Light propagation in (2+1)-dimensional electrodynamics: the case of linear constitutive laws}% 

\author{\'Erico Goulart}
 \email{egoulart@ufsj.edu.br}
\affiliation{Federal University of S\~ao Jo\~ao d'El Rei, C.A.P. Rod.: MG 443, KM 7, CEP-36420-000, Ouro Branco, MG, Brazil
}

\author{Eduardo Bittencourt}
 \email{bittencourt@unifei.edu.br}
\author{Elliton O. S. Brand\~ao}%
 \email{ellitonbrandao@unifei.edu.br}
\affiliation{Federal University of Itajub\' a, BPS Avenue, 1303, Itajub\'a/MG - Brazil
}

\date{\today}

\begin{abstract}
In this paper, we turn our attention to light propagation in three-dimensional electrodynamics. More specifically, we investigate the behavior of light rays in a continuous bi-dimensional hypothetical medium living in a three-dimensional ambient spacetime. Relying on a fully covariant approach, we assume that the medium is endowed with a local and linear response tensor which maps field strengths into excitations. In the geometric optics limit, we then obtain the corresponding Fresnel equation and, using well known results from algebraic geometry, we derive the effective optical metric and indicate possible applications of this formalism in the context of solid-state physics.
\end{abstract}

%\keywords{Suggested keywords}%Use showkeys class option if keyword
                              %display desired
\maketitle

%\tableofcontents

\section{\label{sec:level1} INTRODUCTION}
The study of light propagation in nontrivial media continues to spread new insights into the structure of modern field theories. The main reason behind this relies (we think) on the fertile interplay of areas this task generally requires: optics, premetric electrodynamics, geometric analysis, algebraic geometry and analogue models of gravity \cite{Born,Landau,Post,Hehl1,Hehl2,Hehl3,Hehl4, Schu1, Viss}. A remarkable consequence of these studies is that the dispersion relation of light in a local and linear medium is governed by a quartic homogeneous polynomial in the wave covector, whose coefficients depend in a cubic manner on the medium parameters: electric permittivity, magnetic permeability and magneto-electric cross terms. Although this result was already implicit in the early papers of Bateman and Tamm \cite{Bateman, Tamm}, only in 2002 Rubilar managed to give a rigorous derivation in the most general case \cite{Rubilar}. Since then, this ``Fresnel surface'' has been re-derived by several authors \cite{Lindell,Perlick1,Itin1,Dahl,Favaro} and it is still at the focus of active theoretical and experimental investigations.

The task here is to derive the  dispersion relation and the corresponding effective optical metric for an electromagnetic theory in two spatial dimensions. More specifically, we shall deal with light propagation in a three-dimensional electrodynamics inside material media assuming a local and linear constitutive relation between field strengths and excitations. Mimicking the four-dimensional formalism as far as possible and essentially sticking to the eikonal approximation, we show that: the dispersion relation is determined by a quadratic homogeneous polynomial in the wave covector, whose coefficients depend in a quadratic manner on the medium parameters. In particular, this shows that the dimension reduction completely modifies the algebraic character of the Fresnel surface and so the derivation of the effective optical metric. In other words, light propagation in a genuine three-dimensional theory is not, in general, equivalent to light propagation in a four-dimensional theory restricted to a three-dimensional submanifold. We shall see that the main reason behind this difference relies on the algebraic identities the constitutive tensor must satisfy in three dimensions. It is worth to mention that this claim is not in disagreement with experiments, since it has been shown that the surface parameters characterizing a 2D media are not a two-dimensional limit of the bulk coefficients \cite{You2019} and, therefore, it is expected that the optics is different.

Although there are few theoretical papers on the issues discussed here \cite{Hadamard1923,Maggi2022,Lapidus82,Boito2019}, the scrutiny and applications of two-dimensional media has a wide literature from the experimental perspective, with several technological promises (see \cite{You2019} and references therein). The latter started with the advent of the graphene in 2004 and, since then, the class of known two-dimensional materials enlarged tremendously, with the great interest of the scientists lying on the peculiar optical response of such devices in comparison with their three-dimensional counterparts: a consequence of the mono-layer structure giving a special band distribution for the electrons composing the lattice. In general, the theoretical background for the optical analysis is the four-dimensional Maxwell's theory, where one of the spatial directions is treated as negligible. We shall see that the dimensional reduction, in fact, changes the tensor rank of some constitutive tensors, corroborating the disregard of the coefficients perpendicular to 2D material, but keeping the covariance of the formalism. We leave for experimental physicists the task of deciding which approach better fits the increasing amount of data concerning these materials.

This paper is organized as follows. In Sec.\ \ref{EM2d} we derive the equations of motion for the electrodynamics inside a two-dimensional linear material and study the decomposition of the general constitutive tensor into its irreducible parts. In Sec.\ \ref{disp_rel}, we use the eikonal approximation to get the Fresnel equation, and using the properties of classical adjoint matrices from linear algebra, we find the general expression for the dispersion relation. Next, in Sec.\ \ref{effec_met}, we obtain the effective optical metric for light rays propagating inside 2D materials and, finally, in Sec.\ \ref{appl} we analyze some particular cases for the sake of comparison and completeness.

\section{Electrodynamics in (2+1)}\label{EM2d}

To begin with, we let $(M, g_{ab})$ denote a $3$-dimensional spacetime with signature convention $(-,+,+)$. For the sake of concreteness we assume $M$ to be smooth and globally hyperbolic, but make no further assumptions on the spacelike geometries which foliate the manifold \footnote{Although our considerations are essentially algebraic, global hyperbolicity is necessary for the well-posedness of the Cauchy problem, which we intend to investigate in a future communication.}. We shall be concerned with the electromagnetic field $F_{ab}=-F_{ba}$ in a region of the manifold where a bi-dimensional polarizable medium is present. Therefore, we assume that the macroscopic equations of motion read as (see Appendix \ref{App1} for conventions)
\begin{equation}\label{Maxwell}
P^{ab}_{\phantom a\phantom a ;b}=J^{a},\quad\quad\quad F_{[ab;c]}=0,\quad\quad\quad a,b,c,...=0,1,2.
\end{equation}
Here $P^{ab}=-P^{ba}$ is the excitation tensor, `` ; '' stands for covariant derivative compatible with the metric $g_{ab}$ and $J^{a}$ is the electric current density. We notice that there are only four partial differential equations for a total of six unknowns: the system cannot be solved until a relationship is found between excitations and fields. Consequently, we must supply Eqs.\ (\ref{Maxwell}) with a constitutive law and, for simplicity, we consider local and linear relations of the type
\begin{equation}
P^{ab}=\frac{1}{2}X^{ab}_{\phantom a\phantom a cd}F^{cd},\quad\quad\quad X_{abcd}=-X_{bacd}=-X_{abdc},
\end{equation} 
where the generic double $(2,2)$ form, $X_{abcd}$, is allowed to depend on time and space but not on the electromagnetic field. In three spacetime dimensions, such a constitutive tensor has a total of $9$ independent components and, therefore, has the same degrees of freedom as a generic rank two tensor. For future convenience we define also the contraction maps
\begin{equation}
X^{a}{}_{c}=X^{ab}{}_{cb},\quad\mbox{and}\quad  X=X^{a}_{\phantom a a}.
\end{equation}

As usual, to make contact with standard vector notation, we need to decompose the field strength and the excitation tensor into their corresponding ``electric'' and ``magnetic'' parts. To do so, we start by defining the Hodge dual of $F_{ab}$ as
\begin{equation}
{}^{\star} F^{a}=\frac{1}{2}\varepsilon^{abc}F_{bc},
\end{equation}
with $\varepsilon_{abc}$ denoting the totally antisymmetric Levi-Civita tensor. With this convention, for any timelike, future-directed  and normalized congruence of observers, henceforth denoted by $t^{a}$, we write the decompositions
\begin{eqnarray}\label{F}
&&F_{ab}=(g_{abcd}E^{d}+\varepsilon_{abc}B)t^{c},\quad\quad\quad {}^{\star} F^{a}=(\varepsilon^{a}_{\phantom a cd}E^{d}-\delta^{a}_{\phantom a c}B)t^{c},\\\label{P}
&&P_{ab}=(g_{abcd}D^{d}+\varepsilon_{abc}H)t^{c},\quad\quad\quad {}^{\star} P^{a}=(\varepsilon^{a}_{\phantom a cd}D^{d}-\delta^{a}_{\phantom a c}H)t^{c},
\end{eqnarray}
where $g_{abcd}=g_{ac}g_{bd}-g_{ad}g_{bc}$ is the Kulkarni-Nomizu product of the metric with itself and
\begin{equation}
E^{a}=F^{ab}t_{b},\quad\quad D^{a}=P^{ab}t_{b},\quad\quad B={}^{\star} F^{a}t_{a},\quad\quad H={}^{\star} P^{a}t_{a}.
\end{equation}
We then notice that the electromagnetic field strength and the induction tensor are each constructed from a spacelike vector orthogonal to $t^{a}$ and a pseudo-scalar. Similarly, it can be checked by direct calculation that the constitutive tensor uniquely decomposes as a sum of four independent parts
\begin{equation}\label{Bel}
X_{abcd}=\{-g_{abpq}(g_{cdrs}\mathfrak{A}^{pr}+\varepsilon_{cds}\mathfrak{B}^{p})+\varepsilon_{abq}(g_{cdrs}\mathfrak{C}^{r}+\varepsilon_{cds}\mathfrak{D})\}t^{q}t^{s}.
\end{equation}
The latter is entirely analogous to the so-called Bel decomposition of the Riemann and Weyl tensors in general relativity \cite{Matte, Bel1,Bel2, Senovilla1, Senovilla2} and routine calculations show that 
\begin{equation}\label{tetrad}
\mathfrak{A}_{ac}\equiv -X_{abcd}t^{b}t^{d},\quad\quad \mathfrak{B}_{a}\equiv X^{\star}{}_{abc}\, t^{b}t^{c},\quad\quad \mathfrak{C}_{b}\equiv-{}^{\star} X_{abc}t^{a}t^{c},\quad\quad\mathfrak{D}\equiv {}^{\star} X^{\star}{}_{ab} t^{a}t^{b},
\end{equation}
with the right, left and double Hodge duals defined in the obvious way as
\begin{equation}
X^{\star}{}_{abc}\equiv\frac{1}{2}X_{abpq}\varepsilon^{pq}_{\phantom a\phantom a c},\quad\quad\quad {}^{\star} X_{abc} \equiv\frac{1}{2}X_{pqbc}\varepsilon^{pq}_{\phantom a\phantom a a},\quad\quad {}^{\star} X^{\star}{}_{ab}\equiv\frac{1}{4}X_{pqrs}\varepsilon^{pq}_{\phantom a\phantom a a}\varepsilon^{rs}_{\phantom a\phantom a b}.
\end{equation}
It is clear from the above definitions that 
\begin{equation}
\mathfrak{A}_{ab}t^{b}=0,\quad\quad \mathfrak{A}_{ab}t^{a}=0,\quad\quad\mathfrak{B}_{a}t^{a}=0,\quad\quad \mathfrak{C}_{a}t^{a}=0,
\end{equation}
from which one concludes that the \textit{permittivity matrix} $\mathfrak{A}_{ac}$ carries $4$ independent components, the \textit{magneto-electric terms} $\mathfrak{B}_{a}$ and $\mathfrak{C}_{a}$  carry a total of $4$ and the \textit{inverse permeability} $\mathfrak{D}$ carries the remaining $1$. This is in sharp contrast with the $4$-dimensional case, where each of these terms would be described by a generic $3\times 3$ matrix \footnote{We refer the reader to \cite{Dell1970,Birss}, where a wealth of details concerning the so-called constitutive matrix may be consulted in the four-dimensional case.}. Finally, combining Eqs. (\ref{F}), (\ref{P}) and (\ref{Bel}), there follow 
\begin{equation}
D^{a}=\mathfrak{A}^{a}_{\phantom a b}E^{b}+\mathfrak{B}^{a}B,\quad\quad\quad H=\mathfrak{C}_{r}E^{r}+\mathfrak{D}B.
\end{equation}
It should be clear from this construction that the \textit{constitutive tetrad} $\{\mathfrak{A}_{ab},\mathfrak{B}_{a},\mathfrak{C}_{a},\mathfrak{D}\}$ is an observer-dependent set and, therefore, depends implicitly on the choice of the auxiliary vector field $t^{a}$. In principle, we could write-down the equations of motion explicitly in terms of the above quantities, but this will not be necessary for our subsequent analysis.

\section{Dispersion relation}\label{disp_rel}
What can be said about light propagation inside the medium? This leads us to the corresponding dispersion relation. In general, the latter is obtained either using Hadamard's method of weak discontinuities \cite{Hadamard} or the eikonal approximation \cite{Perlick1, Visser1}, which we now apply. Roughly speaking, we assume an approximate wavy solution to Eqs. (\ref{Maxwell}) of the form
\begin{equation}
F_{ab}\approx f_{ab}(x)e^{i\Theta(x)},
\end{equation}
with $f_{ab}(x)$ a slowly varying amplitude and $\Theta(x)$ a rapidly varying phase. In this eikonal approximation we neglect gradients in the amplitude and retain only the gradients of the phase $\partial_{a}\Theta\equiv k_{a}$. This is enough to achieve the regime of geometrical optics, where the notion of light rays are well defined. A simple calculation shows that the Bianchi identity in Eqs.\ (\ref{Maxwell}) gives 
\begin{equation}
f_{ab}k_{c}+f_{ca}k_{b}+f_{bc}k_{a}=0,
\end{equation}
from which one concludes that the amplitude of the electromagnetic disturbance reduces to a simple 2-form, which may be written as
\begin{equation}
f_{ab}=k_{a}a_{b}-k_{b}a_{a},
\end{equation}
with $a_{a}$ denoting the polarization 1-form. Applying the latter to the first equation in Eqs.\ (\ref{Maxwell}) gives the nontrivial algebraic condition
\begin{equation}\label{main}
(X^{ambn}k_{m}k_{n})a_{b}=0,
\end{equation}
which is the building block of the dispersion relation: it implies an algebraic constraint which must be fulfilled by the characteristic covectors $k_{a}$ in order to obtain a physically meaningful solution. It is worth mentioning that, up to now, the eikonal approximation has lead us to exactly the same equations as in the four-dimensional analogue. It should also be remarked that a corresponding expression can be obtained from the metric-free approach \cite{Hehl1}, as described in Appendix \ref{App2}.

In order to investigate the algebraic implications of Eq.\ (\ref{main}) in more details, we proceed very much in the same way as in \cite{Itin1, Dahl, Favaro}. Since our considerations here are essentially algebraic, we shall fix a point $x$ on the manifold and consider the map
\begin{equation}\label{ymap}
Y: T_{x}^{*}M\rightarrow\mbox{Mat}_{3\times 3}(\mathbb{R}),\quad\quad q_{m}\mapsto Y^{ab}(q)\equiv X^{ambn}q_{m}q_{n}.
\end{equation}
An important property of this map is that every covector in the domain produces a matrix which automatically annihilates the corresponding covector. In other words, we have
\begin{equation}\label{algcond}
Y^{ab}(q)q_{b}=Y^{ba}(q)q_{b}=0.
\end{equation}
Roughly speaking, this means that the image of $T_{x}^{*}M$ in the nine-dimensional space $\mbox{Mat}_{3\times 3}(\mathbb{R})$ is not arbitrary, but rather belongs to the eight-dimensional determinantal variety, defined by $\mbox{rk}\ Y^{ab}(q)\leq 2$. When combined with the Cayley-Hamilton theorem, this fact guarantees that
\begin{equation}
Y^{a}_{\phantom a c}Z^{c}_{\phantom a b}=Z^{a}_{\phantom a c}Y^{c}_{\phantom a b}=0,
\end{equation}
where
\begin{eqnarray}\label{adjointmatrix}
Z^{a}_{\phantom a b}&\equiv&\frac{1}{2}\delta^{apq}_{\phantom a\phantom a\phantom a brs}Y^{r}_{\phantom a p}Y^{s}_{\phantom a q}=Y^{a}_{\phantom a c}Y^{c}_{\phantom a b}-\sigma_{1}Y^{a}_{\phantom a b}+\sigma_{2}\delta^{a}_{\phantom a b},
\end{eqnarray}
is the classical adjoint tensor with the following $k$-th elementary symmetric polynomials 
\begin{equation}
\sigma_{1}\equiv Y^{p}_{\phantom a p},\quad\quad\quad \sigma_{2}\equiv\frac{1}{2}(Y^{p}_{\phantom a p}Y^{q}_{\phantom a q}-Y^{p}_{\phantom a q}Y^{q}_{\phantom a p}),\quad\quad\quad \sigma_{3}\equiv \mbox{det}(Y^{a}_{\phantom a b})=0.
\end{equation}
Clearly, the classical adjoint is a quadratic combination of the constitutive tensor $X^{abcd}$ and a quartic combination of the covector $q_{m}$. Essentially, it is here that the three-dimensional case departures from the four-dimensional one: the rank of a given element of $\mbox{Mat}_{3\times 3}(\mathbb{R})$ will be directly related to the structure of its adjoint, and a well known result of linear algebra adapted to three dimensions states that
\begin{itemize}
\item{when $\mbox{rk}\ Y^{ab}(q)=2$, then $\mbox{rk}\ Z^{ab}(q)=1$;}
\item{when $\mbox{rk}\ Y^{ab}(q)=1$, then $Z^{ab}(q)=0$.}
\end{itemize}
In what follows, in order to distinguish between the two types of covectors, we shall identify $q_{a}$ with $k_{a}$ when the the second condition is fulfilled. Accordingly, the matrix $Y^{a}_{\phantom a b}(k)$ will belong to a subset of the determinantal variety of dimension five.  

Let us suppose first that $\mbox{rk}\ Y^{a}_{\phantom a b}(q)=2$, for some nonzero covector $q_{a}$. In this case, simple algebraic manipulations using Eq.\ (\ref{algcond}) show that the adjoint must have a trivial dyadic structure of the form
\begin{equation}\label{Zed}
Z_{ab}= H(x,q)q_{a}q_{b},\quad\quad\quad\mbox{where}\quad\quad\quad H(x,q)\sim \hat{g}^{ab}(x)q_{a}q_{b}.
\end{equation} 
Here the homogeneous quadratic function $H(x,q)$ is characterized by a second order contra-variant tensor $\hat{g}^{ab}$, henceforth called \textit{effective optical metric} \footnote{Strictly speaking, it is the inverse tensor $\hat{g}_{ab}$ that formally defines a metric tensor, as long as $\hat{g}^{ab}$ is nondegenerate. Besides, $\hat{g}_{ab}$ shall not be confused with $g_{ac}g_{bd}\hat{g}^{cd}$.}, which is a quadratic combination of the constitutive tensor $X^{abcd}$. Interestingly, due to continuity arguments, in order to obtain the restricted case of $\mbox{rk}\ Y^{a}_{\phantom a b}(k)=1$, we need to impose
\begin{equation}\label{disprel}
\hat{g}^{ab}(x)k_{a}k_{b}=0.
\end{equation}
Up to an arbitrary conformal factor, this equation has the form of the dispersion relation we are looking for. In other words, Eq.\ (\ref{main}) will admit nontrivial solutions if and only if the corresponding wave covectors coincide with the vanishing set of the quadratic polynomial defined above. This is a direct consequence of the rank-nullity theorem and the fact that $\mbox{ker}\ Y^{ab}(k)=\mbox{span}(k_{b}, a_{b})$ in our case. 

\section{effective optical metric}\label{effec_met}
In this section, we shall calculate the explicit form of the function $H(x,q)$. To do so, we first consider the following lemma, whose proof involves straightforward manipulations of three-dimensional Levi-Civita tensors and generalized Kronecker deltas (see e.g. \cite{Herdeiro})

\begin{lemma}
Let $A^{\{\Omega\}}_{\phantom a\phantom a\phantom a ab}$ and $B_{ab}{}^{\{\Upsilon\}}$ be two tensors with $\{\Omega\}$ and $\{\Upsilon\}$ schematically denoting a generic number of indices, plus a pair of skew indices $ab$. Then, in three dimensions:
\begin{equation}
(A^{\star}{}^{\{\Omega\}a})({}^{\star} B_{b\{\Upsilon\}})=A^{\{\Omega\}}_{\phantom a\phantom a\phantom a mb}B^{ma}_{\phantom a\phantom a\phantom a\{\Upsilon\}}-\frac{1}{2}A^{\{\Omega\}}_{\phantom a\phantom a\phantom a mn}B^{mn}_{\phantom a\phantom a\phantom a\{\Upsilon\}}\delta^{a}_{\phantom a b}.
\end{equation}
\end{lemma}
\noindent The main feature of the identity is that it involves a term containing no contraction on the left hand side, a term with a single contraction and a term with two contractions. Applying the latter to the constitutive tensor and its Hodge duals gives
\begin{cor}
Putting $A^{\star}{}^{\{\Omega\}a}\rightarrow X^{\star}{}^{i_{1}i_{2}i_{3}}$ and ${}^{\star} B_{b\{\Upsilon\}}\rightarrow {}^{\star} X_{j_{1}j_{2}j_{3}}$ one gets the identity
\begin{equation}
(X^{\star}{}^{i_{1}i_{2}i_{3}})({}^{\star} X_{j_{1}j_{2}j_{3}})=X^{i_{1}i_{2}}_{\phantom a\phantom a\phantom a k_{1}j_{1}}X^{k_{1}i_{3}}_{\phantom a\phantom a\phantom a j_{2}j_{3}}-\frac{1}{2}X^{i_{1}i_{2}}_{\phantom a\phantom a\phantom a k_{1}k_{2}}X^{k_{1}k_{2}}_{\phantom a\phantom a\phantom a\phantom a j_{2}j_{3}}\delta^{i_{3}}_{\phantom a j_{1}}.
\end{equation}
\end{cor}

\begin{cor}\label{cor2}
Putting $A^{\star}{}^{\{\Omega\}a}\rightarrow {}^{\star} X^{\star}{}^{i_{1}i_{2}}$ and ${}^{\star} B_{b\{\Upsilon\}}\rightarrow {}^{\star} X^{\star}{}_{j_{1}j_{2}}$ one gets the identity
\begin{equation}
({}^{\star} X^{\star}{}^{i_{1}i_{2}})({}^{\star} X^{\star}{}_{j_{1}j_{2}})=({}^{\star} X^{i_{1}}{}_{k_{1}j_{1}})(X^{\star}{}^{k_{1}i_{2}}{}_{j_{2}})-\frac{1}{2}({}^{\star} X^{i_{1}}{}_{ k_{1}k_{2}})(X^{\star}{}^{k_{1}k_{2}}{}_{ j_{2}})\delta^{i_{2}}{}_{j_{1}}.
\end{equation}
\end{cor}

In order to compute the homogeneous quadratic function $H(x,q)$ using the above identities, we start by realizing that the trace of the adjoint matrix, as defined in Eq.\ (\ref{adjointmatrix}), is proportional to the second elementary symmetric polynomial. Combining this fact with the trivial dyadic structure of the adjoint Eq. (\ref{Zed}), one obtains
\begin{eqnarray}\label{did}
H(x,q)q^{2}&=&\frac{1}{2}(Y^{k_{1}k_{2}}Y_{k_{2}k_{1}}-Y^{k_{1}}_{\phantom a\phantom a k_{1}}Y^{k_{2}}_{\phantom a\phantom a k_{2}})=\frac{1}{2}( X^{k_{1}i_{2}k_{2}i_{1}}X_{k_{2}j_{2}k_{1}j_{1}}-X^{i_{1}i_{2}}X_{j_{1}j_{2}})q_{i_{1}}q_{i_{2}}q^{j_{1}}q^{j_{2}},
\end{eqnarray}
with $q^{2}=q^{k}q_{k}$, for conciseness. The problem of finding $H(x,q)$ thus reduces to showing that: 

\begin{theo}
The homogeneous fourth order polynomial on the right hand side of Eq.\ (\ref{did}) factorizes as a product of two homogeneous quadratic polynomials.
\end{theo}

\proof
Since this algebra is somehow cumbersome, we present the calculations in their full details. 

First, recalling that $X^{i_{1}}_{\phantom a j_{1}}=X^{i_{1}k_{1}}_{\phantom a\phantom a\phantom a j_{1}k_{1}}$ and $X=X^{k_{1}}_{\phantom a\ k_{1}}$, one easily proves the $(2+1)$-dimensional analogue of the so-called Ruse-Lanczos identity
\begin{equation}\label{RL}
{}^{\star} X^{\star}{}^{i_{1}j_{1}}=X^{j_{1}i_{1}}-\frac{1}{2}X g^{j_{1}i_{1}}.
\end{equation}
Now, contracting $i_{2}$ with $j_{1}$ in Corollary 2 gives
\begin{equation}\label{corollarya}
({}^{\star} X^{\star}{}^{i_{1}k_{1}})({}^{\star} X^{\star}{}_{k_{1}j_{2}})=-\frac{1}{2}({}^{\star} X^{i_{1}}{}_{k_{1}k_{2}})(X^{\star}{}^{k_{1}k_{2}}{}_{j_{2}}),
\end{equation}
and reinserting this in the original equation, yields
\begin{equation}\label{corollaryb}
({}^{\star} X^{i_{1}}{}_{k_{1}j_{1}})(X^{\star}{}^{k_{1}i_{2}}{}_{j_{2}})=({}^{\star} X^{\star}{}^{i_{1}i_{2}})({}^{\star} X^{\star}{}_{j_{1}j_{2}})-({}^{\star} X^{\star}{}^{i_{1}k_{1}})({}^{\star} X^{\star}{}_{k_{1}j_{2}})\delta^{i_{2}}{}_{j_{1}}.
\end{equation}
Contracting $i_{1}$ with $j_{2}$ in Corollary 1 and rearranging the terms gives
\begin{equation}\label{corollaryc}
({}^{\star} X^{i_{1}}_{\phantom a\phantom a k_{1}j_{1}})(X^{\star}{}^{k_{1}i_{2}}_{\phantom a\phantom a\phantom a\ j_{2}})=X^{k_{1}i_{2}k_{2}i_{1}}X_{k_{2}j_{2}k_{1}j_{1}}+\frac{1}{2}X^{i_{2}}_{\phantom a\phantom a k_{1}k_{2}k_{3}}X^{k_{2}k_{3}k_{1}}_{\phantom a\phantom a\phantom a\phantom a\phantom a\phantom a j_{1}}\delta^{i_{1}}_{\phantom a j_{2}}.
\end{equation}
Thus, combining Eq.\ (\ref{corollaryb}) with Eq.\ (\ref{corollaryc}), one gets
\begin{equation}\label{corollaryd}
 X^{k_{1}i_{2}k_{2}i_{1}}X_{k_{2}j_{2}k_{1}j_{1}}=({}^{\star} X^{\star}{}^{i_{1}i_{2}})({}^{\star} X^{\star}{}_{j_{1}j_{2}})-({}^{\star} X^{\star}{}^{i_{1}k_{1}})({}^{\star} X^{\star}{}_{k_{1}j_{2}})\delta^{i_{2}}{}_{j_{1}}-\frac{1}{2}X^{i_{2}}_{\phantom a\phantom a k_{1}k_{2}k_{3}}X^{k_{2}k_{3}k_{1}}_{\phantom a\phantom a\phantom a\phantom a\phantom a\phantom a j_{1}}\delta^{i_{1}}_{\phantom a j_{2}}.
\end{equation}
Multiplying Eq. (\ref{corollaryd}) by $q_{i_{1}}q_{i_{2}}$ and $q^{j_{1}}q^{j_{2}}$ and using Eq. (\ref{RL}) reveals that
\begin{equation}
Y^{k_{1}}_{\phantom a\phantom a k_{2}}Y^{k_{2}}_{\phantom a\phantom a k_{1}}-Y^{k_{1}}_{\phantom a\phantom a k_{1}}Y^{k_{2}}_{\phantom a\phantom a k_{2}}=-q^{2}\left(X^{i_{1}}_{\phantom a\phantom a k_{1}}X^{k_{1}}_{\phantom a\phantom a j_{2}}+\frac{1}{2}X^{i_{1}}_{\phantom a\phantom a k_{1}k_{2}k_{3}}X^{k_{2}k_{3}k_{1}}_{\phantom a\phantom a\phantom a\phantom a\phantom a\ j_{2}}\right)q_{i_{1}}q^{i_{2}}
\end{equation}
and, finally, contracting $i_{2}$ with $j_{1}$ and $i_{3}$ with $j_{3}$ in Corollary 1 gives the desired result
\begin{equation}\label{final}
Y^{k_{1}}_{\phantom a\phantom a k_{2}}Y^{k_{2}}_{\phantom a\phantom a k_{1}}-Y^{k_{1}}_{\phantom a\phantom a k_{1}}Y^{k_{2}}_{\phantom a\phantom a k_{2}}=q^{2}[(X^{\star}{}^{i_{1}}{}_{k_{1}k_{2}})({}^{\star} X^{k_{1}k_{2}}{}_{j_{2}})]q_{i_{1}}q^{j_{2}}.
\end{equation}
Eq.\ (\ref{final}) shows that the quartic multivariate polynomial in Eq.\ (\ref{did}) is indeed proportional to the squared norm of the covector, as expected. \qed

Furthermore, since Eq. (\ref{did}) is valid for all covectors in $T_{x}^{*}M$, there follows
\begin{equation}
H(x,q)=\frac{1}{2}[(X^{\star}{}^{i_{1}}{}_{k_{1}k_{2}})({}^{\star} X^{k_{1}k_{2}i_{2}})]q_{i_{1}}q_{i_{2}}.
\end{equation}
Comparing this relation with Eq. (\ref{Zed}) we then read off the effective optical metric as
\begin{equation}\label{effmet}
 \hat{g}^{ i_{1}i_{2}}(x)=-\frac{1}{2}{}^{\star} X^{k_{1}k_{2}(i_{1}}X^{\star}{}^{i_{2})}{}_{k_{1}k_{2}}
\end{equation}
with the minus sign chosen for convenience. We notice that Eq. (\ref{effmet}) is the 3-dimensional analogue of the celebrated Tamm-Rubilar tensor, which in its turn is cubic in the constitutive tensor of the medium and governs light propagation in $4$-dimensional electrodynamics. 

So far, we have derived the effective optical metric treating the constitutive tensor $X^{abcd}$ as a whole. However, in practical situations, one is more often concerned with its smaller projected pieces: the constitutive tetrad. In order to compute Eq.\ (\ref{effmet}) explicitly in terms of these pieces, we recall Eq.\ (\ref{Bel}) to write the right and left Hodge duals, respectively, as
\begin{eqnarray*}
X^{\star}{}{}_{abc}&=&\{g_{abpq}(-\varepsilon_{crs}\mathfrak{A}^{pr}+g_{cs}\mathfrak{B}^{p})+\varepsilon_{abq}(\varepsilon_{crs}\mathfrak{C}^{r}-g_{cs}\mathfrak{D})\}t^{q}t^{s},\\\\
{}^{\star} X_{bcd}&=&\{-\varepsilon_{bpq}(g_{cdrs}\mathfrak{A}^{pr}+\varepsilon_{cds}\mathfrak{B}^{p})-g_{bq}(g_{cdrs}\mathfrak{C}^{r}+\varepsilon_{cds}\mathfrak{D})\}t^{q}t^{s}.
\end{eqnarray*}
After a lengthy but straightforward calculation, Eq.\ (\ref{effmet}) then becomes
\begin{equation}\label{final1}
\hat{g}^{ab}(x)=\mathfrak{D}\mathfrak{A}^{(ab)}+\frac{1}{2}(\mathfrak{A}^{p}_{\phantom a q}\mathfrak{A}^{q}_{\phantom a p}-\mathfrak{A}^{p}_{\phantom a p}\mathfrak{A}^{q}_{\phantom a q})t^{a}t^{b}-\mathfrak{B}^{(a}\mathfrak{C}^{b)}+\varepsilon_{pqr}(\mathfrak{A}^{p(b}t^{a)}\mathfrak{B}^{q}-t^{(a}\mathfrak{A}^{b)p}\mathfrak{C}^{q})t^{r}.
\end{equation}
In the particular case of an ordinary material medium without magneto-electric terms, we get
\begin{equation}\label{final2}
\hat{g}^{ab}(x)=\mathfrak{D}\mathfrak{A}^{(ab)}+\frac{1}{2}(\mathfrak{A}^{p}_{\phantom a q}\mathfrak{A}^{q}_{\phantom a p}-\mathfrak{A}^{p}_{\phantom a p}\mathfrak{A}^{q}_{\phantom a q})t^{a}t^{b}.
\end{equation}

\section{Applications}\label{appl}

In this section, we briefly investigate some interesting consequences of Eqs.\ (\ref{final1}) and (\ref{final2}). More precisely, we consider particular constitutive laws for the cases of vacuum, isotropic, anisotropic, linear magneto-electric and pure skewonic media. To do so, we start by writing the projector tensor orthogonal to the observer as
\begin{equation}
h^{ab}\equiv g^{ab}+t^{a}t^{b},   
\end{equation}
which satisfies the relations
\begin{equation}\label{projector}
h_{ab}=h_{(ab)},\quad\quad\quad h_{ab}t^{b}=0,\quad\quad\quad h^{a}_{\phantom a c}h^{c}_{\phantom a b}=h^{a}_{\phantom a b},\quad\quad\quad h^{a}_{\phantom a a}=2.    
\end{equation}
We notice also that a generic constitutive tensor in three dimensions is irreducibly decomposed into symmetric and antisymmetric (skewonic) parts as
\begin{equation}
X_{abcd}={\phantom a}^{(s)}X_{abcd}+{\phantom a}^{(a)}X_{abcd}    
\end{equation}
where
\begin{eqnarray*}
&&{\phantom a}^{(s)}X_{abcd}\equiv \frac{1}{2}(X_{abcd}+X_{cdab})\quad\rightarrow\quad {\phantom a}^{(s)}X_{abcd}=+{\phantom a}^{(s)}X_{cdab},\\
&&{\phantom a}^{(a)}X_{abcd}\equiv \frac{1}{2}(X_{abcd}-X_{cdab})\quad\rightarrow\quad{\phantom a}^{(a)}X_{abcd}=-{\phantom a}^{(a)}X_{cdab}.\\
\end{eqnarray*}
When combined with Eq. (\ref{Bel}), the latter gives the following splitting of the constitutive tetrad\\
\begin{equation}\label{consttet}
\{\mathfrak{A}_{ab},\mathfrak{B}_{a},\mathfrak{C}_{a},\mathfrak{D}\}=\{{}^{(s)}\mathfrak{A}_{ab},{}^{(s)}\mathfrak{B}_{a},-{}^{(s)}\mathfrak{B}_{a},{}^{(s)}\mathfrak{D}\}\oplus \{{}^{(a)}\mathfrak{A}_{ab},{}^{(a)}\mathfrak{B}_{a},{}^{(a)}\mathfrak{B}_{a},0\},
\end{equation}\\
where ${}^{(s)}\mathfrak{A}_{ab}$ is symmetric and ${}^{(a)}\mathfrak{A}_{ab}$ is antisymmetric. Therefore, the symmetric part has a total of $6$ independent components whereas the antisymmetric part has a total of $3$. This is in sharp contrast with the four-dimensional case, where the former would have $21$ and the latter $15$. Furthermore, we recall that an additional totally antisymmetric part (the axion) is also allowed in the four-dimensional decomposition. That the latter does not appear in our analysis is a direct consequence of the algebraic identity $X_{[abcd]}=0$, valid for all rank four tensors in three dimensions. 

\subsection{Vacuum medium}
This case is characterized by the simple symmetric constitutive tensor $X_{abcd}=g_{abcd}$. Using Eqs.\ (\ref{tetrad}) with the splitting Eq.\ (\ref{consttet}), the nonvanishing elements of the constitutive tetrad read as
\begin{equation}
{}^{(s)}\mathfrak{A}_{ab}=h_{ab},\quad\quad\quad{}^{(s)}\mathfrak{D}=1,
\end{equation}
which is absent of all possible magneto-electric cross terms and has a symmetric permittivity matrix. Using Eq. (\ref{final2}) with (\ref{projector}), one obtains the effective optical metric and its inverse as
\begin{equation}
 \hat{g}^{ab}=g^{ab}\quad\quad \leftrightarrow\quad\quad \hat{g}_{ab}=g_{ab},
\end{equation}
which recovers the well known fact that in vacuum, the dispersion relation is governed by the background spacetime metric itself. The next two examples consist of generalizations of this result for two types of symmetric medium without magneto-electric parts.

\subsection{Isotropic medium}
This case is governed by a constitutive tetrad of the type
    \begin{equation}
{}^{(s)}\mathfrak{A}_{ab}=\varepsilon h_{ab},\quad\quad\quad{}^{(s)}\mathfrak{D}=\mu^{-1},
\end{equation}
where the electric permittivity $\varepsilon$ and the (inverse) magnetic permeability $\mu^{-1}$ are arbitrary functions of position in spacetime. Again, using Eq.\ (\ref{final2}) with (\ref{projector}), one obtains (up to a conformal factor)
\begin{equation}
 \hat{g}^{ab}=g^{ab}+(1-\mu\epsilon)t^{a}t^{b}\quad\quad \leftrightarrow\quad\quad \hat{g}_{ab}=g_{ab}+\left(1-\frac{1}{\mu\varepsilon}\right)t_{a}t_{b}
\end{equation}
This is precisely the metric obtained by Gordon and Pham Mau Quan in the classic references \cite{Gordon, Pham}, but now restricted to two spatial dimensions. In order to ensure that the propagation of light rays are well defined in the effective spacetime, the determinant of $\hat{g}_{ab}$ must be negative definite. In a local frame such that $g_{ab}(x)=\eta_{ab}$ and $t^{a}=\delta^{a}_{\phantom a 0}$, a direct calculation gives
\begin{equation}
\det(\hat{g}_{ab})=-1/\mu\varepsilon,
\end{equation}
that is always negative if the product $\mu\varepsilon$ is positive. This condition encompasses most dielectric materials found in nature. That the lower dimensional case presented here reproduces the same behavior of the four-dimensional one is a direct consequence of the simple algebraic symmetries assumed for the constitutive tensor.

\subsection{Anisotropic medium}
In order to introduce an anisotropic behavior (as in the case of crystals) at a spacetime point, it suffices to consider a symmetric electric permittivity matrix whose principal values do not coincide. In other words, we consider a constitutive tetrad of the form
    \begin{equation}
{}^{(s)}\mathfrak{A}_{ab}=\begin{pmatrix}
    0 & 0 & 0\\
    0 & \varepsilon_{1} & 0\\
    0 & 0 & \varepsilon_{2}\end{pmatrix},\quad\quad\quad{}^{(s)}\mathfrak{D}=\mu^{-1},
\end{equation}
where the constitutive parameters $\varepsilon_{1}$, $\varepsilon_{2}$ and $\mu^{-1}$ are all allowed to depend on position and $t^{a}=\delta^{a}_{\phantom a 0}$, for simplicity. Using Eq. (\ref{final2}), there follows (up to a conformal factor)
\begin{equation}
\hat{g}^{ab}=\begin{pmatrix}
    -\varepsilon_{1}\varepsilon_{2} & 0 & 0\\
    0 & \varepsilon_{1}/\mu & 0\\
    0 & 0 & \varepsilon_{2}/\mu \end{pmatrix}\quad\quad\leftrightarrow\quad\quad
\hat{g}_{ab}=\begin{pmatrix}
    -1/\varepsilon_{1}\varepsilon_{2} & 0 & 0\\
    0 & \mu/\varepsilon_{1} & 0\\
    0 & 0 &\mu/\varepsilon_{2} \end{pmatrix}
    \end{equation}
    The reader is invited to consult \cite{Born, Landau} for similar results in the four-dimensional case. Now, the determinant of the effective optical metric is
\begin{equation}
\label{det_anis_opt_met}
\det(\hat{g}_{ab})=-\mu^2/\varepsilon_1^2\varepsilon_2^2,
\end{equation}
which is always negative for non-vanishing parameters. However, in order to obtain the physically meaningful signature, we need to impose either the positivity or negativity of all constitutive parameters.

\subsection{Magneto-electric medium}
We now consider a particular type of anisotropic medium endowed with generic magneto-electric terms. Hence, in a local frame such that $g_{ab}(x)=\eta_{ab}$ and $t^{a}=\delta^{a}_{\phantom a 0}$, we assume a constitutive tetrad as follows
    \begin{equation}
{}^{(s)}\mathfrak{A}_{ab}=\begin{pmatrix}
    0&0&0\\0&\varepsilon_{1}&0\\0&0&\varepsilon_{2}
    \end{pmatrix},\quad\quad \mathfrak{B}_{a}=\begin{pmatrix}
    0&\mathfrak{B}_{x}&\mathfrak{B}_{y}\end{pmatrix},\quad\quad \mathfrak{C}_{a}=\begin{pmatrix}
    0&\mathfrak{C}_{x}&\mathfrak{C}_{y}
    \end{pmatrix},\quad\quad{}^{(s)}\mathfrak{D}=\mu^{-1},
\end{equation}
where the dielectric parameters $\varepsilon_1$, $\varepsilon_2$ and $\mu$ as well as the magneto-electric ones $\mathfrak{B}_{x}$, $\mathfrak{B}_{y}$, $\mathfrak{C}_{x}$ and $\mathfrak{C}_{y}$ are arbitrary functions of the spacetime coordinates. Here the effective optical metric, given by Eq.\ (\ref{final1}) can be written (up to a conformal factor) as
\begin{equation}
\hat{g}^{ab}=\begin{pmatrix}
     -\varepsilon_{1}\,\varepsilon_{2}&\frac{1}{2}\,(\mathfrak{B}_{y}-\mathfrak{C}_{y})\,\varepsilon_{1}&-\frac{1}{2}\,(\mathfrak{B}_{x}-\mathfrak{C}_{x})\,\varepsilon_{2} \\\frac{1}{2}\,(\mathfrak{B}_{y}-\mathfrak{C}_{y})\,\varepsilon_{1}&\varepsilon_{1}/\mu-\mathfrak{B}_{x}\,\mathfrak{C}_{x}&-\frac{1}{2}\,(\mathfrak{B}_{x}\,\mathfrak{C}_{y}+\mathfrak{B}_{y}\,\mathfrak{C}_{x})\\-\frac{1}{2}\,(\mathfrak{B}_{x}-\mathfrak{C}_{x})\,\varepsilon_{2} &-\frac{1}{2}\,(\mathfrak{B}_{x}\,\mathfrak{C}_{y}+\mathfrak{B}_{y}\,\mathfrak{C}_{x})&\varepsilon_{2}/\mu-\mathfrak{B}_{y}\,\mathfrak{C}_{y} 
     \end{pmatrix}.
\end{equation}
After some manipulations, one can show that the determinant of this metric can be put in the form
\begin{equation}
\det(\hat{g}^{ab})=-\frac{1}{4}\left(\frac{\varepsilon_{1}\,\varepsilon_{2}}{\mu}-\varepsilon_{1}\mathfrak{B}_{y}\mathfrak{C}_{y}-\varepsilon_{2}\mathfrak{B}_{x}\mathfrak{C}_{x}\right)\left[4\frac{\varepsilon_{1}\,\varepsilon_{2}}{\mu}+\varepsilon_{1}(\mathfrak{B}_{y}-\mathfrak{C}_{y})^2+\varepsilon_{2}(\mathfrak{B}_{x}-\mathfrak{C}_{x})^2\right].
\end{equation}    
From the latter, one realizes that Lorentzian signature cannot be guaranteed without further assumptions on the magneto-electric terms. For instance, if the products of magneto-electric coefficients are positive and large enough to exceed the first term in parenthesis, then the effective optical metric has an Euclidean signature and there is no good propagation. On the hand, if those terms are negative, then $\hat{g}^{ab}$ has a Lorentzian signature and we have propagation again. Finally, for sufficiently weak magneto-electric media, where the product of any two magneto-electric coefficients can be neglected, the determinant $\hat{g}^{ab}$ reduces to Eq.\ (\ref{det_anis_opt_met}).

\subsection{Pure skewonic medium}
The last case we analyze here corresponds to a  hypothetical medium whose constitutive tensor $X_{abcd}$ contains only its antisymmetric part with respect to the change of skew indices. In a local frame such that $g_{ab}(x)=\eta_{ab}$ and $t^{a}=\delta^{a}_{\phantom a 0}$, we assume a constitutive tetrad as follows
\begin{equation}
    {}^{(a)}\mathfrak{A}_{ab}=\begin{pmatrix}
    0&0&0\\0&0&-\varepsilon\\0&\varepsilon&0
    \end{pmatrix},\quad\quad {}^{(a)}\mathfrak{B}_{a}=\begin{pmatrix}0&\mathfrak{B}_{x}&\mathfrak{B}_{y}\end{pmatrix},\quad\quad {}^{(a)}\mathfrak{C}_{a}=\begin{pmatrix}0&\mathfrak{B}_{x}&\mathfrak{B}_{y}\end{pmatrix},\quad\quad{}^{(a)}\mathfrak{D}=0.
\end{equation}
Consequently, Eq.\ (\ref{final1}) gives for the effective optical metric the following expression (up to a conformal factor)
\begin{equation}
    \hat{g}^{ab}= \begin{pmatrix}-\varepsilon^{2}&-\varepsilon\mathfrak{B}_{x}&-\varepsilon\mathfrak{B}_{y}\\-\varepsilon\mathfrak{B}_{x}&-\mathfrak{B}_{x}{}^{2}&-\mathfrak{B}_{x}\mathfrak{B}_{y}\\-\varepsilon\mathfrak{B}_{y}&-\mathfrak{B}_{x}\mathfrak{B}_{y}&-\mathfrak{B}_{y}{}^{2}\end{pmatrix}.
\end{equation} 
Interestingly, the rank of the effective optical metric is one and, therefore, there is no room for hyperbolicity in pure skewonic media in two spatial dimensions. This result is entirely consistent with Itin`s claim in four dimensions that the skewon part alone does not provide a non-trivial dispersion relation. Thus, in a three-dimensional spacetime as well, the skewon can serve only as a supplement to the principal (symmetric) part (see \cite{Skew} for details).

\section{Conclusion}
With the help of the eikonal approximation, the algebraic properties of the constitutive tensor $X_{abcd}$ of an electromagnetic theory inside a (2+1)-dimensional medium led us to the dispersion relation and the effective optical metric. In particular, we show that such relation can be written as a quadratic homogeneous polynomial in the wave covector whose coefficients also depend quadratically on the medium parameters. Then, we studied cases of interest, for instance, isotropic/anisotropic dielectrics and magneto-electric media, emphasizing the necessary conditions for a well-defined propagation of light rays.

The recent and increasing interest of experimentalists and technologists on the optical features of two-dimensional medias have expanded this area faster than its theoretical counterpart, leaving some conceptual questions behind, for instance, the absence of a covariant description of manifestly 2D phenomena. On the other hand, we expect that the approach described along our text may shed some light towards an explanation of the optical phenomena measured in laboratory, particularly, the effective number of degrees of freedom of the medium since the elements of the constitutive tetrad have different tensor rank in comparison to the same set in the (3+1)-dimensional formalism.

Beyond the limits of geometric optics, it is well established that a genuine covariant 2D electromagnetic field is useful in describing the constitutive law associated to the Quantum Hall effect \cite{Hehl1}, because of the topological character of the Hall current. It is possible to set a linear phenomenological relation between the Hall current $j^{\mu}$ and the Hodge dual of the Faraday tensor ${}^{\star}F^{\mu}$ of the form $j^{\mu}=\sigma_{H}{}^{\star}F^{\mu}$, where $\sigma_{H}$ is the Hall resistance. Setting a congruence of observers, it yields the correct two-dimensional relations $\vec j=\sigma_H\vec E$ and $\rho=-\sigma_H B$, where $\vec j$ and $\rho$ are the 2D Hall current and the charge density, respectively. On the other hand, in the last decades, the development of photonic \cite{Raghu2008} and spin \cite{Wu2018} analogues of quantum Hall effects have given rise to broad interest in topological phenomena described by 3D electromagnetism and quantum mechanics.

It is also worth to be noticed that a two-dimensional electromagnetism in fact admits two possible formulations, based upon the \textit{method of descent} proposed by Hadamard \cite{Hadamard1923,Maggi2022}. It means that the electromagnetic field could be represented either by a 2-form $F_{ab}$ (as we proceed here) or by a 1-form $F_a$, leading to nonequivalent formulations. However, adding the extra assumption of planar invariance of the Lorentz force, it is easy to show that the approach the 2-form $F_{ab}$ formulation is favored over the other. Furthermore, it is equivalent to the three-dimensional Maxwell's theory restricted to a plane where the fields, charges and currents are independent of the direction perpendicular to the plane of symmetry \cite{Lapidus82,Boito2019}. 
Ultimately, for further investigation, we shall address in separate the case of a nonlinear constitutive relation \cite{paperii}, where the discussion in terms of phase and group velocities and polarization seems very enlightening. We also intend to study other mathematical aspects of this theory, for example, the characteristics of propagation for degenerate effective optical metric and the interplay between the causal structure and the covariant hyperbolizations as discussed in \cite{symm1,symm2}.

\acknowledgments
EOSB would like to thank CNPq for the financial support (Grant number 134395/2021-2).

\appendix
\section{Conventions}\label{App1}

Throughout, all lower-case Latin indices take their values in the set $0,1,2$ and the velocity of light in vacuum is normalized to unity ($c\equiv 1$). The Levi-Civita tensors are defined by
\begin{equation}
\varepsilon_{abc}=\sqrt{-g}[abc],\quad\quad\quad \varepsilon^{abc}=-\frac{1}{\sqrt{-g}}[abc],   
\end{equation}
where $g\equiv\mbox{det}(g_{ab})$ in any coordinate system and $[abc]$ is the totally antisymmetric symbol, with $[012]=+1$. The generalized Kronecker delta of order $k$ is defined by the multilinear determinant
\begin{equation}
\delta^{a_{1}\ldots a_{k}}{}_{b_{1}\ldots b_{k}}\equiv \mbox{det}\left|\begin{array}{cccc}
\delta^{a_{1}}{}_{b_{1}}& \delta^{a_{1}}{}_{b_{2}}&\cdots &\delta^{a_{1}}{}_{b_{k}}\\
\delta^{a_{2}}{}_{b_{1}}&\delta^{a_{2}}{}_{b_{2}}&\cdots &\delta^{a_{2}}{}_{b_{k}}\\
\vdots &\vdots &\ddots &\vdots \\
\delta^{a_{k}}{}_{b_{1}}&\delta^{a_{k}}{}_{b_{2}}&\cdots &\delta^{a_{k}}{}_{b_{k}}\\
\end{array}\right|,
\end{equation}
and there follow the fundamental identities
\begin{equation}
\varepsilon^{abc}\varepsilon_{pqr}=-\delta^{abc}{}_{pqr},\quad\quad \varepsilon^{abr}\varepsilon_{pqr}=-\delta^{ab}{}_{pq},\quad\quad \varepsilon^{aqr}\varepsilon_{pqr}=-2\delta^{a}{}_{p},\quad\quad \varepsilon^{pqr}\varepsilon_{pqr}=-6.
\end{equation}
As usual, total antisymmetrization of a tensor is defined as
\begin{equation}
T_{[a_{1}...a_{k}]}\equiv\frac{1}{k!}\delta^{b_{1}\ldots b_{k}}{}_{a_{1}\ldots a_{k}}T_{b_{1}\ldots b_{k}},    
\end{equation}
and for all $k\geq 4$, there follows $T_{[a_{1}...a_{k}]}=0$. In particular, the latter shows that an arbitrary constitutive tensor in three dimensions contain no axionic counterpart.

\nocite{*}

\section{Pre-metric approach}\label{App2}

From a theoretical standpoint it is often convenient to formulate electromagnetism on a bare manifold that need not carry a metric or a connection. This tradition dates back at least to the works of Kottler, Cartan, van Dantzig and Schr\"odinger and is at the roots of the pre-metric approaches developed in \cite{Post, Hehl1}. For the sake of comparison with the metric formalism adopted in the paper we briefly review here some essential features of the metric-free approach. To start with, we consider a three-dimensional bare manifold $M$ and treat the electromagnetic field strength $F_{ab}(x)$ as a closed untwisted (even) $2$-form. In order to construct consistent equations of motion for the field, we are allowed to use only objects permitted by the differential structure of the manifold i.e., tensors, densities, partial derivatives and related quantities. Following Refs.\ \cite{Hehl1,Perlick1}, we write Maxwell equations in (2+1)-dimensions as
\begin{equation}\label{metric-free}
F_{[ab,c]}=0,\quad\quad\quad\quad H_{[a,b]}=-J_{ab}.  
\end{equation}
Here $H_{a}$ is the electromagnetic excitation and $J_{ab}$ is the electric current and the minus sign is chosen for convenience. We notice that, for this particular dimension, all quantities above carry exactly the same number of degrees of freedom. Under coordinate transformations, it is assumed that these fields change according to
\begin{eqnarray*}
\tilde{F}_{ab}=\frac{\partial x^{c}}{\partial \tilde{x}^{a}}\frac{\partial x^{d}}{\partial \tilde{x}^{b}}F_{cd},\quad\quad\quad \tilde{H}_{a}=\mbox{sgn}\left(\mbox{det}\ \frac{\partial x}{\partial \tilde{x}}\right)\frac{\partial x^{b}}{\partial \tilde{x}^{a}}H_{b},\quad\quad\quad \tilde{J}_{ab}=\mbox{sgn}\left(\mbox{det}\ \frac{\partial x}{\partial \tilde{x}}\right)\frac{\partial x^{c}}{\partial \tilde{x}^{a}}\frac{\partial x^{d}}{\partial \tilde{x}^{b}}J_{cd}.
\end{eqnarray*}
Hence, $H_{a}$ transforms as a twisted (odd) $1$-form whereas $J_{ab}$ transforms as a twisted (odd) $2$-form. In particular, this distinction turns out to be important in the realm of optics in questions related to chirality. It is clear that Eqs.\ (\ref{metric-free}) are metric-free and that they must be compatible with
\begin{equation}
J_{[ab,c]}=0,
\end{equation}
which is nothing but the conservation law for the electric charge.

In order to make the theory consistent from the point of view of partial differential evolution equations, we must supply it with a ``spacetime'' local relation of the form
\begin{equation}\label{kappa}
F_{ab}\ \mapsto\ H_{a}=\frac{1}{2}\kappa_{a}^{\phantom a bc}F_{bc},
\end{equation}
where the mixed twisted quantity $\kappa_{a}^{\phantom a bc}=-\kappa_{a}^{\phantom a cb}$ carries a total of $9$ independent components, is allowed to depend on position and provides a linear map between the space of untwisted $2$-forms to the space of twisted $1$-forms. Defining the Levi-Civita tensor densities of weights $+1$ and $-1$ respectively by $\check{\varepsilon}^{abc}$ and $\check{\varepsilon}_{abc}$, %one has the relation $\check{\varepsilon}^{abc}\check{\varepsilon}_{def}=\delta^{abc}{}_{def}$ and the latter may be used to
we write down Eqs.\ (\ref{metric-free}) in the form
\begin{equation}\label{Maxchi}
F_{[ab,c]}=0,\quad\quad\quad\quad(\check{H}^{ab})_{,b}=\check{J}^{a},
\end{equation}
where
\begin{equation}
\check{H}^{ab}\equiv \check{\varepsilon}^{abc}H_{c},\quad\quad\quad \check{J}^{a}\equiv\frac{1}{2}\check{\varepsilon}^{abc}J_{bc},\quad\quad\quad \check{\varepsilon}^{abc}\check{\varepsilon}_{def}=\delta^{abc}{}_{def}.    
\end{equation}
A closer inspection of Eq.\ (\ref{kappa}) then shows that the constitutive law may be alternatively written as
\begin{equation}\label{chi}
\check{H}^{ab}=\frac{1}{2}\chi^{abcd}F_{cd},\quad\quad\mbox{with}\quad\quad\quad \chi^{abcd}\equiv \check{\varepsilon}^{abm}\kappa_{m}^{\phantom a cd}.
\end{equation}
The $9$ functions $\chi^{abcd}(x)$ contain the same information as $\kappa_{a}^{\phantom a bc}(x)$ and has the following symmetries
\begin{equation}
\chi^{abcd}=-\chi^{bacd}=-\chi^{abdc}.
\end{equation}
Due to the properties of the Levi-Civita symbol, the latter qualifies as an untwisted tensor density of weight $+1$ (see reference \cite{Hehl1}, pag. 247). It is clear that Maxwell's equations in their metric-dependent form Eqs.\ (\ref{Maxwell}) are a particular instance of Eq.\ (\ref{Maxchi}). Indeed, the former may be obtained via the identifications 
\begin{equation}
\check{H}^{ab}\ \rightarrow\ \sqrt{-g}P^{ab},\quad\quad\quad \check{J}^{a}\ \rightarrow\ \sqrt{-g}J^{a}.
\end{equation}
The important point here is that Hadamard's method of discontinuities may be straightforwardly applied to Eqs.\ (\ref{Maxchi}) thus yielding a dispersion relation of the form Eq.\ (\ref{effmet}) with just minor algebraic adaptations.

\bibliography{ref}

\end{document}